\documentclass[print,twocolumn]{revtex4}
\usepackage{amsmath}
\usepackage{pifont}
\usepackage{amssymb}
\usepackage{tipa}
\usepackage{amsfonts}
\usepackage{mathrsfs}
\usepackage{graphicx}
\usepackage{dcolumn}
\usepackage{bm}
\usepackage{txfonts}
\usepackage[dvips]{color}
\usepackage[colorlinks,linkcolor=red,citecolor=blue]{hyperref}

\begin{document}

\title{Anisotropic magneto-optical absorption and linear dichroism in two-dimensional semi$-$Dirac electron systems}
\author{Xiaoying Zhou}
\email{xiaoyingzhou@hunnu.edu.cn}
\author{Wang Chen}
\author{Xianzhe Zhu}
\affiliation{Department of Physics, Key Laboratory for Low-Dimensional Structures and
Quantum Manipulation (Ministry of Education), Hunan Normal University, Changsha 410081, China}

\begin{abstract}
We present a theoretical study on the Landau levels (LLs) and
magneto-optical absorption of a two-dimensional semi-Dirac electron system under a
perpendicular magnetic field. Based on an effective \textbf{\emph{k$\cdot$p}}
Hamiltonian, we find that the LLs are proportional to the
two-thirds power law of the magnetic field and level index, which can be understood as a
hybridization of the LL of Schr\"{o}dinger and Dirac electrons with new features.
With the help of Kubo formula, we find the selection
rule for interband (intraband) magneto-optical transition is anisotropic
(isotropic). Specifically, the selection rules for interband magneto-optical transitions are
$\Delta n$=0, $\pm2$ ($\pm2$, $\pm4$) for linearly polarized light along the
linear (parabolic) dispersion direction, while the selection rules for the
intraband transition are $\Delta n$=$\pm1$, $\pm3$ regardless of the
polarization direction of the light. Further, the
magneto-optical conductivity for interband (intraband) transition excited by linearly polarized light along the
linear dispersion direction is two (one) orders of magnitude larger than
that along the parabolic dispersion direction. This anisotropic
magneto-optical absorption spectrum clearly reflects the structure of the
LLs, and results in a strong linear dichroism. Importantly,
a perfect linear dichroism with magnetic-field tunable wavelength can be
achieved by using the interband transition between the two lowest LLs, i.e, from $E_{v0}$ to
$E_{c0}$. Our results shed light on the magneto-optical property of the two dimensional semi-Dirac electron systems and pave the way to design magnetically
controlled optical devices.
\end{abstract}

\maketitle



\section{Introduction}

In the past decade, the study on the Dirac-Weyl fermions in condensed
matter systems has attracted intensive attention on account of both rich
physics therein and promising applications \cite{Neto,Armitage}. Two-dimensional (2D) semi-Dirac material is a new kind of highly
anisotropic electron system, of which the low energy dispersion is linear along one direction and parabolic along
the perpendicular direction \cite{Dietl,Montambaux1,Banerjee1}. Owing to the unique dispersion, a semi-Dirac
material has the properties of both Dirac materials and conventional semimetals or semiconductors \cite{Dietl,Montambaux1,Banerjee1}. Various systems are predicted to host
2D semi-Dirac electrons such as the anisotropic strain modulated
graphene \cite{Dietl,Montambaux1}, the multi-layer
$\mathrm{(TiO_2)_n/(VO_2)_m}$ nanostructures \cite{Pardo1,Banerjee1,Huang} and the
strained or electric field modulated few-layer black phosphorus \cite{Baik,Wang1,Liu,Ghosh,Doh}.
Recently, semi-Dirac electron has been observed experimentally in
potassium-doped and strained few-layer phosphorene \cite{Kim,Makino}.

Although the low energy dispersion of a semi-Dirac material is a
hybridization of that in Dirac materials and conventional semimetals, it still exhibits unique features which cannot be fully
understood by combing the existed results of the two typical materials. Those features include
the unusual Landau levels (LLs) \cite{Dietl,Banerjee1,Montambaux1}, the optical
conductivity \cite{Jang,Carbotte}, the anisotropic plasmon \cite{Banerjee2},
the Fano factor in ballistic transport \cite{Zhai2}, the power-law decay indexes in
the quasi-particle interference spectrum \cite{XiaoyingZhou}, and so on. In
particular, the LLs of 2D semi-Dirac electron system are proportional to the two-thirds
power law of the magnetic field and level index, i.e., $E_n\propto[(n+1/2)B%
]^{2/3}$ \cite{Montambaux1}, which has been verified by the
magneto-transport experiment \cite{Makino}. This is different from the
linear dependence on the magnetic field and the level index in conventional
semimetals or semiconductors \cite{Stern,Klitzing} and the square root power
dependence in pure Dirac materials \cite{KSNovoselov,Zhangyuanbo,Sadowski}.
The LL structure is an important fundamental issue for electron material
systems because it dominates the magneto-properties, such as the
quantum Hall effect, magneto-optical and -resonance features  of the
material \cite{Stern,Klitzing,KSNovoselov,Zhangyuanbo,Sadowski}. In turn,
magneto-measurement is also a powerful tool to detect the structure of the
LLs \cite{Klitzing,KSNovoselov,Zhangyuanbo,Sadowski,Jiang,PengCheng,Hanaguri,likaiLi,Crassee}.
Further, the band parameters such as the effective masses and the Fermi
velocity can be extracted from the measured LL spectrum, which has been
successfully applied in graphene \cite{KSNovoselov,Zhangyuanbo,Sadowski,Jiang,Crassee}, the
surface states of three-dimensional topological insulators \cite{PengCheng,Hanaguri},
and 2D black phosphorus \cite{likaiLi}. To date, there are already several
theoretical works on the LLs of the semi-Dirac electron system \cite{Dietl,Banerjee1,Montambaux1}
and also a magneto-transport measurement on it \cite{Makino}. However, the magneto-optical
property of 2D semi-Dirac electron system still remains elusive.

In this work, we theoretically investigate the LLs and magneto-optical
properties of a 2D semi-Dirac electron system under a perpendicular magnetic field. By
fitting the formula giving by the Sommerfeld quantization with the numerical
calculations, we find that the LLs are proportional to the 2/3
power of the magnetic field and the level index, which is different from that in the
conventional semi-metals or semiconductors \cite{Stern,Klitzing} and the massless
Dirac materials \cite{KSNovoselov,Zhangyuanbo,Sadowski}. There is a band gap in the LL spectrum, and the LL spacings decrease with the increase of the level index.
With the help of Kubo formula, we evaluate the longitudinal
magneto-optical conductivity as functions of the photon energy. We find
the selection rule for interband (intraband) magneto-optical transitions
is anisotropic (isotropic). The selection rules for interband transitions are $\Delta n$=0,$\pm2$ ($\pm2,\pm 4$) for linearly polarized
light along the linear (parabolic) dispersion direction, while the
selection rules for intraband transitions are $\Delta n$=$\pm1$, $\pm3$
regardless of the polarization direction of the light. For interband (intraband)
transitions, the magneto-optical conductivity excited by the linearly polarized light
along the linear dispersion direction is hundreds (dozens) times larger
than that along the parabolic direction dispersion. The anisotropic
magneto-optical absorption spectrum clearly reflects the LL structure
and results in a strong linear dichroism. Importantly, the interband
transition between the two lowest-LLs results in a perfect linear dichroism
with magnetic field tunable wavelength, which is useful to design
magneto-optical devices.

The rest of the paper is organized as follows. In Sec. II, we introduce the
calculation of LLs and magneto-optical properties based on Kubo formula. In
Sec. III, we present the numerical results and discussions. Finally, we
summarize our results in Sec. IV.

\section{Landau levels and Magneto-Optical transitions}

\subsection{Landau levels}

The low energy effective Hamiltonian of a 2D semi-Dirac electron system is \cite{Dietl,Baik}
\begin{equation}
H=\frac{p_{y}^{2}}{2m^{\ast }}\sigma _{x}+v_{F}p_{x}\sigma _{y},
\end{equation}%
where $\sigma_{x}$ and $\sigma_{y}$ are the Pauli matrices, $\mathbf{p}$=$(\hbar k_{x},\hbar k_{y})$
the momentum, $m^{\ast}$ the effective mass and $v_{F}$ the Fermi velocity.
Typically, in potassium doped few-layer black
phosphorus, the two parameters are \cite{Baik} $v_{F}=3\times 10^{5}$ m/s
and $m^{\ast}=1.42$ $m_{0}$, where $m_{0}$ is the free electron mass. The
eigenvalue of Hamiltonian (1) is
\begin{equation}
E_{s}=s\sqrt{p_{y}^{4}/4m^{\ast 2}+v_{F}^{2}p_{x}^{2}},
\end{equation}%
where the sign $s$=$+/-$ stands for the conduction/valence band. Eq. (2) indicates that
the low energy state of a 2D semi-Dirac electron system is highly anisotropic, of which
the dispersion is linear (parabolic) in the $k_x(k_y$)-direction.
When a perpendicular magnetic field $\mathbf{B}$= (0,0,B) is applied,
performing the Peierls substitution $\mathbf{p}\rightarrow \mathbf{\pi =p+eA},$ we have the
commutator $\left[ \pi _{x},\pi _{y}\right] =-ieB\hbar $. Hence,
the upper and lower operators can be defined as
\begin{equation}
\hat{a}=\frac{l_{B}}{\sqrt{2}\hbar }\left( \pi _{x}-i\pi _{y}\right) ,\hat{a}%
^{\dag }=\frac{l_{B}}{\sqrt{2}\hbar }\left( \pi _{x}+i\pi _{y}\right) ,
\end{equation}%
where $l_{B}$=$\sqrt{\hbar /eB}$ is the magnetic length. Therefore, Hamiltonian
(1) can be rewritten as
\begin{equation}
H=-\frac{\hbar ^{2}}{4m^{\ast }l_{B}^{2}}(\hat{a}^{\dag }-\hat{a})^{2}\sigma
_{x}+\frac{\hbar v_{F}}{\sqrt{2}l_{B}}(\hat{a}^{\dag }+\hat{a})\sigma _{y}.
\end{equation}%
This Hamiltonian can not be solved analytically because the lower and upper
operators couple all the LLs together. Fortunately, it can be solved numerically by taking the
eigenvectors of the number operator $\hat{n}$=$\hat{a}^{\dag}\hat{a}$ as
basis functions. In this basis, the wave function of the system can be
written as
\begin{equation}
\psi =\sum_{m=0}^{M}\left(
\begin{array}{c}
u_{m} \\
v_{m}%
\end{array}%
\right) |m\rangle ,
\end{equation}%
where $u_{m}$ and $v_{m}$ are the linear superposition coefficients, and $M$
is the total number of basis functions. Then, we can diagonalize Hamiltonian
(4) numerically in a truncated Hilbert space and obtain the eigenvalues as
well as the eigenvectors. In Landau gauge $\bm{A}$=$(-By,0,0)$, the basis
function $|m\rangle $ is $\langle r|m\rangle$=$\varphi (x,y)$=$\frac{%
e^{ik_{x}x}}{\sqrt{L_{x}}}\phi _{m}\left( y-y_{0}\right) $, where $\phi _{m}(.)$ is the eigenvector of one dimensional harmonic oscillator, and $y_{0}$=$k_{x}l_{B}^{2}$ is the cyclotron center.

\begin{figure}[tbp]
\center
\includegraphics[bb=22 30 668 556, width=0.49\textwidth]{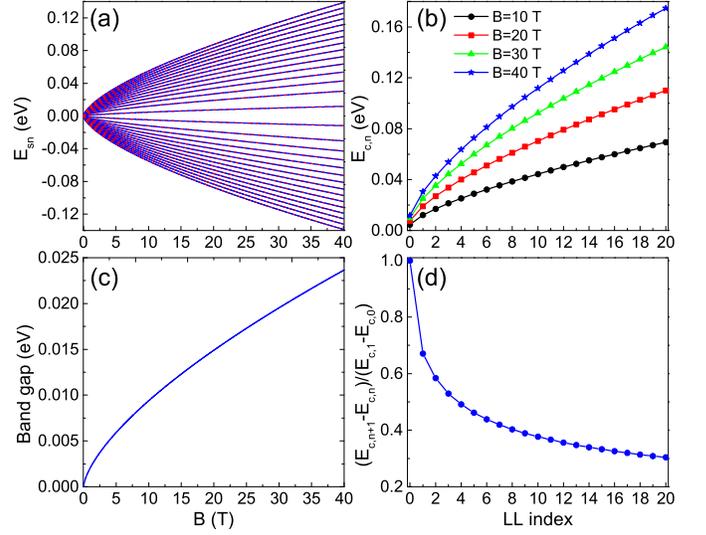} %
\renewcommand{\figurename}{FIG.}
\caption{(a) LLs as a function of magnetic field for the lowest fifteen
sub-bands, where the blue solid (red dashed) lines indicate the numerical
(analytical) results. (b) LLs in the conduction band varying with level
index under different magnetic fields. (c) Band gap of the LL spectrum
versus magnetic field. (d) LL spacings in the conduction band (blue line) in
unit of the first one ($E_{c,1}-E_{c,0}$) as a function of level index under
magnetic field $B=30$ T.}
\end{figure}

Another way to obtain the eigenvalues of Hamiltonian (4) is to use a
semiclassical argument, e.g., the Sommerfeld quantization \cite{Dietl}. The
formula of LL spectrum is given as
\begin{equation}
E_{s,n}=sg(n)[\frac{\hbar v_{F}e}{\sqrt{m^{\ast }}}(n+\frac{1}{2}%
)B]^{2/3},n=0,1,2,3,\cdots.
\end{equation}%
To determine the function $g(n)$, we need to fit Eq. (6) with the numerical
data. In our work, we find $g(0)$=0.9454, $g(1)$=1.1668, and $g(n)$=1.1723
for $n$$\geqq$2. At this point, it is interesting to compare the LLs of
2D semi-Dirac electron systems with those of Schr\"{o}dinger electrons in conventional
semi-metal or semiconductors, and Dirac electrons in graphene. The results are
summarized as
\begin{equation}
E_{n}=\left\{
\begin{array}{c}
\frac{\hbar eB}{m^{\ast }}(n+\frac{1}{2}):\text{Schr\"{o}dinger electrons}
\\
\text{sgn}\left( n\right) v_{F}\sqrt{2e\hbar B\left\vert n\right\vert }:%
\text{Dirac electrons} \\
\text{sgn}(n)[\sqrt{\frac{\hbar eB}{m^{\ast }}}v_{F}\sqrt{e\hbar B}(|n|+\frac{1}{2}%
)]^{2/3}:\text{Semi-Dirac electrons.}%
\end{array}%
\right.
\end{equation}%
Obviously, the LLs of the three kinds of two dimensional electron gas are different from
each other. Further, the LL spacings for $n$$\geqq$2 are
\begin{eqnarray}
\Delta E_n &=&E_{s,n+1}-E_{s,n}  \notag \\
&=&\frac{2\left( n+1\right) \epsilon _{0}}{(n+\frac{3}{2})^{4/3}+(n+\frac{1}{%
2})^{4/3}+(n+\frac{3}{2})^{2/3}(n+\frac{1}{2})^{2/3}},
\end{eqnarray}%
where $\epsilon_{0}=\left(\hbar v_{F}eB/\sqrt{m^{\ast }}\right)^{2/3}.$

Figure 1(a) plots the LLs as a function of magnetic field for the lowest
fifteen sub-bands. From Fig. 1(a), we find that the LLs (the red lines)
given by Sommerfeld quantization [see Eq. (6)] perfectly reproduce the
numerical results (blue lines) under any magnetic field. This proves that
the LLs of 2D semi-Dirac electron system are proportional to the 2/3 power law of
the magnetic field. Moreover, the LLs also show $2/3$ power law dependence
on the level index under different magnetic fields [see Fig. 1(b)]. This
$2/3$ power dependence on the magnetic field and level index is
different from that in conventional semi-metals or semiconductors \cite%
{Stern,Klitzing} or Dirac materials \cite{KSNovoselov,Zhangyuanbo,Sadowski}
. Although semi-Dirac electrons are realized in potassium-doped few-layer black phosphorus \cite{Kim}, the LLs are already quite different from those of pristine black phosphorus \cite{xyzhou1,xyzhou2,xyzhou3,yjjiang}, indicating that they have become different electron systems. In contrast to the gapless LLs of Dirac materials \cite{KSNovoselov,Zhangyuanbo,Sadowski}, there is a band gap in the LL spectrum due to the zero-point energy of the
harmonic oscillator, which is more similar to that in conventional semiconductors \cite{Stern,Klitzing}. The band gap is $2E_{c,0}$ which increases with the $2/3 $ power law of
the magnetic field [see Fig. 1(c)]. The stronger the magnetic field, the
larger the band gap. Further, Fig. 1(d) shows the LL spacings in the conduction band
in unit of the first one ($E_{c,1}- E_{c,0}$) as a function of level index
under magnetic field $B$=30 T. From Fig. 1(d), we find that the LL spacings
decrease with increasing level index, which can also be inferred from Eq.
(8). In other words, the higher the level index, the smaller the LL spacing. For high level
index limitation ($n$$\rightarrow$$\infty$), the LL spacing vanishes.

Unlike the highly anisotropic dispersion at zero magnetic field [see
Eq. (2)], the LLs of a 2D semi-Dirac electron system are independent of the wave vectors, and
seem to be isotropic.  However, the anisotropy of the LLs can be revealed from
the wave functions. Figs. 2(a) and 2(b) plot the spatial probability
distributions in different Landau gauges of the first and second LL, respectively. The probability
distributions are calculated by using the finite difference method \cite{Smith} which is
not presented here. As plotted in Fig. 2, we find that the
probability distribution exhibits strong anisotropy. It decays much faster
along the $y$-direction (red lines) than that along the $x$-direction (blue
lines). This means that electrons are more localized in the $y$-direction
due to the larger effective mass along this direction. The highly
anisotropic probability distribution (wave function) plays important role in the magneto-optical absorption spectrum of 2D semi-Dirac electron system
as shall be discussed later.

To conclude this subsection, the LL spectrum of a 2D semi-Dirac electron can be understood as a
hybridization of that of the Schr\"{o}dinger and Dirac electron but with
new features. In particular, the band gap arising from the zero-point energy
is inherited from the Schr\"{o}dinger electron, while the decreasing LL
spacing is inherited from the Dirac electron. However, the 2/3 power law dependence
on the magnetic field and level index as well as the highly anisotropic wave
function are not embedded in the LL spectrum of Schr\"{o}dinger and Dirac
electron systems.
\begin{figure}[t]
\center
\includegraphics[bb=39 2 797 592, width=0.48\textwidth]{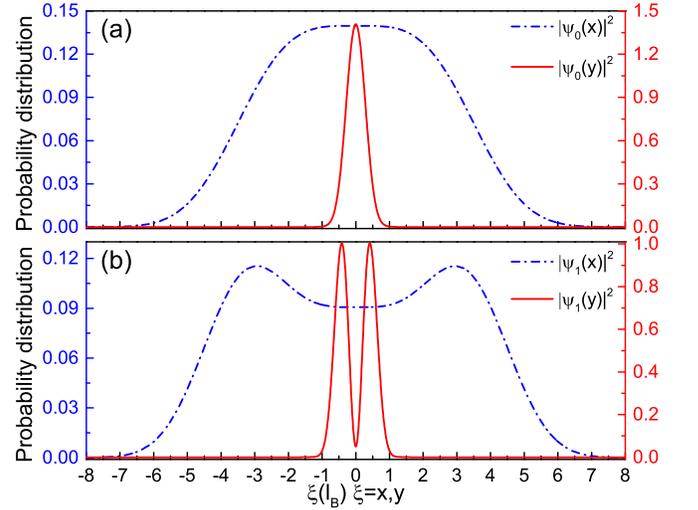} %
\renewcommand{\figurename}{FIG.}
\caption{(a)/(b) The spatial probability distributions of the first/second
LLs in the conduction when choose different gauges for the vector potential $%
\mathbf{A}$. The blue/red lines represent the probability distribution
along the $x/y$-direction.}
\end{figure}

\begin{figure*}[tbp]
\center 
\includegraphics[width=0.9\textwidth]{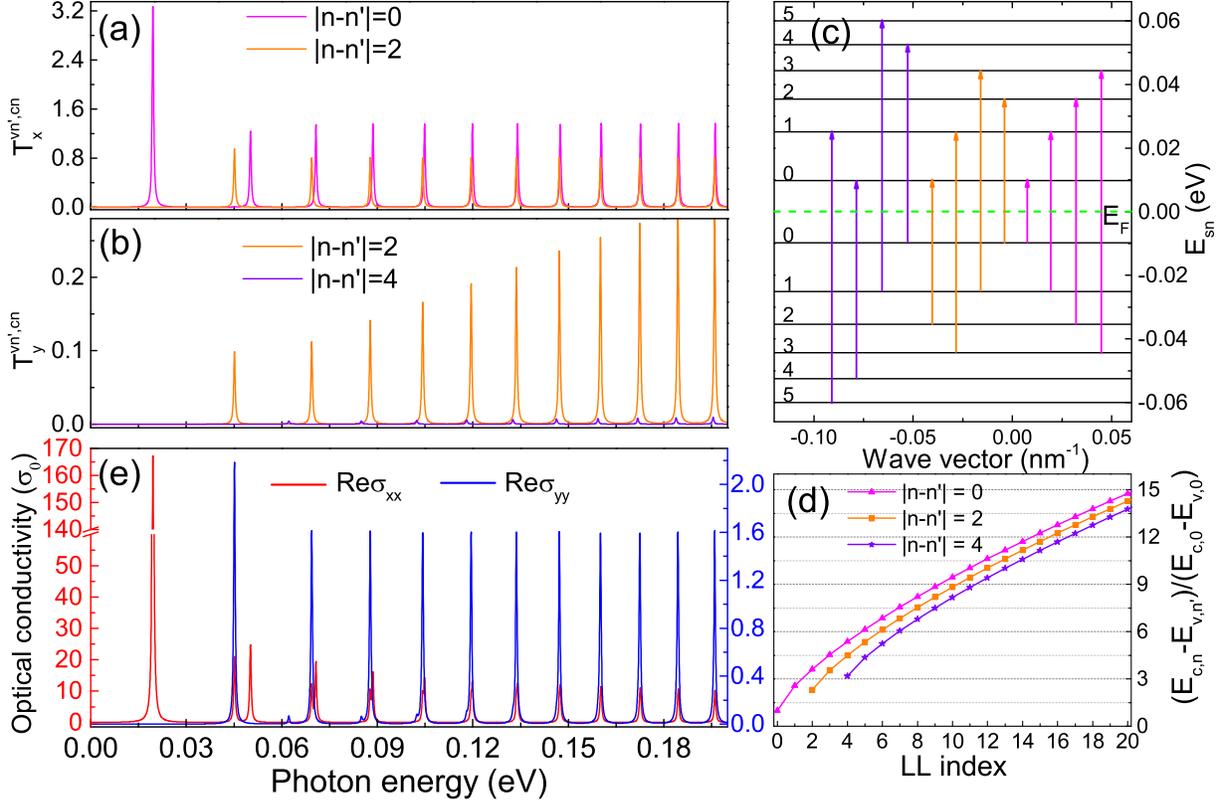} \renewcommand{%
\figurename}{Fig.}
\caption{(a)/(b) Nonzero elements of the interband transition rate as a
function of photon energy for linearly polarized light along the $x$/$y$-direction
with magnetic field $B=30$ T. (c) Schematic illustration of the
interband magneto-optical transition rules, where the magenta, orange and
purple arrows denote the interband transitions of $\Delta n=|n-n^{\prime }|=0, \pm 2,\pm 4$,
respectively. (d) The photon energies of the three kinds of interband
transitions with $|n-n^{\prime }|=0, \pm 2,\pm 4$ in unit of the band gap as
a function of level index. (e) The real part of the longitudinal
magneto-optical conductivity (in units of $\protect\sigma_0=2e^2/h$) for
interband transitions as a function of photon energy with magnetic field $%
B=30$ T. }
\end{figure*}

\subsection{Magneto-optical properties}

Within the linear response theory, the dynamical conductivity can be written
as \cite{TAndo1,Ando,Lizhou,Tabert}
\begin{equation}
\sigma _{\mu \nu }(\omega )=\frac{\hbar e^{2}}{iS_{0}}\sum_{\xi \neq \xi
^{\prime }}\frac{[f(E_{\xi })-f(E_{\xi ^{\prime }})]\langle \xi |v_{\mu
}|\xi ^{\prime }\rangle \langle \xi ^{\prime }|v_{\nu }|\xi \rangle }{%
(E_{\xi }-E_{\xi ^{\prime }})(E_{\xi }-E_{\xi ^{\prime }}+\hbar \omega
+i\Gamma _{\xi })},
\end{equation}%
where $\hbar \omega $ is the photon energy, $S_{0}$$=$$L_{x}L_{y}$ the
sample area with the size $L_{x}$ ($L_{y}$) in the $x$ ($y$)-direction, $%
|\zeta \rangle $$=$$|s,n,k_{x}\rangle $ the wavefunction expressed in Eq. (5), $f(E_{\xi })$$=$$%
[e^{(E_{\zeta }-E_{F})/k_{B}T}+1]^{-1}$ the Fermi-Dirac distribution
function with Boltzman constant $k_{B}$ and temperature $T$. The sum runs
over all states $|\xi \rangle $$=$$|s,n,k_{x}\rangle $ and $|\xi ^{\prime
}\rangle $$=$$|s^{\prime },n^{\prime },k_{x}^{\prime }\rangle $ with $\xi $$%
\neq $$\xi ^{\prime }$. Meanwhile, $\Gamma _{\xi }\propto \sqrt{B}$ accounts for
the LL broadening induced by the disorder effects \cite{TAndo1,Esfarjani}. In
the simplest approximation, we can assume the broadenings are the same for
each LL. The velocity matrix operators $v_{x/y}$=$\partial H/\partial p_{x/y}$ are
\begin{equation}
v_{x}=v_{F}\sigma _{y},v_{y}=-iv_{0}(\hat{a}^{\dag }-\hat{a})\sigma _{x},
\end{equation}%
where $v_{0}$=$\hbar /\sqrt{2}l_{B}m^{\ast }$. It is worth noting that the velocity operators are anisotropic which will result in a highly anisotropic magneto-optical absorption spectrum. Under moderate magnetic fields, the absorption for linearly polarized light along the $x$-direction is stronger than that along the $y$-direction because of $v_F>v_0$. Meanwhile, we note that $v_x$ ($v_y$) is independent (dependent) on the upper and lower operators, which means the magneto-optical transition selection rules may also be anisotropic. By using the wavefunction in
Eq. (5), the transition matrix elements of the velocity matrices are
\begin{eqnarray}
X_{n^{\prime },n}^{s^{\prime },s} &=&\left\langle s^{\prime },n^{\prime
},k_{x}^{\prime }\left\vert v_{x}\right\vert s,n,k_{x}\right\rangle   \notag \\
&=&\sum_{m^{\prime },m}^{M}iv_{F}(-u_{m^{\prime }}^{n^{\prime },s^{\prime
}\ast }v_{m}^{n,s}+v_{m^{\prime }}^{n^{\prime },s^{\prime }\ast
}u_{m}^{n,s})\delta _{m^{\prime },m},  \notag \\
Y_{n,n^{\prime }}^{s,s^{\prime }} &=&\left\langle s,n,k_{x}\left\vert
v_{y}\right\vert s^{\prime },n^{\prime },k_{x}^{\prime }\right\rangle  \notag
\\
&=&\sum_{m,m^{\prime }}^{M}-iv_{0}(u_{m}^{n,s\ast }v_{m^{\prime
}}^{n^{\prime },s^{\prime }}+v_{m}^{n,s\ast }u_{m^{\prime }}^{n^{\prime
},s^{\prime }})  \notag \\
&&\times \left( -\sqrt{m}\delta _{m^{\prime },m-1}+\sqrt{m+1}\delta
_{m^{\prime },m+1}\right) .
\end{eqnarray}%
By Fermi's golden rule \cite{Zettili,Sari}, the transition rate from the $n$-th LL in the $s$ band to the $%
n^{\prime }$-th one in the $s^{\prime }$ band for linearly polarized light
along the $x$-direction is
\begin{equation}
T_{x}^{s^{\prime }n^{\prime },sn}=\frac{2\pi }{\hbar }\left( \frac{\hbar }{%
l_{B}}\left\vert X_{n,n^{\prime }}^{s,s^{\prime }}\right\vert \right)
^{2}\delta \left( E_{s^{\prime }n^{\prime }}-E_{sn}\pm \hbar \omega \right)
f(E_{s^{\prime }n^{\prime }})\left[ 1-f(E_{sn})\right] ,
\end{equation}%
while $T_{y}^{s^{\prime }n^{\prime },sn}$ is similar to $T_{x}^{s^{\prime
}n^{\prime },sn}$. Hence, the normal of the matrix elements $|X_{n^{\prime
},n}^{s^{\prime },s}|^{2}$ and $|Y_{n^{\prime },n}^{s^{\prime },s}|^{2}$
directly determine the magneto-optical transition selection rules. A zero matrix element
represents a forbidden transition. Although the transition matrix elements
[see Eq. (11)] can not be obtained analytically, we can still obtain the
magneto-optical transition selection rules by numerically checking all the
matrix elements of the transition rate one by one. With the velocity matrix
elements in Eq. (11), one can evaluate the magneto-optical
conductivity for linearly polarized light directly. Substituting Eq. (11)
into Eq. (9) and making the replacement $\sum_{k_{x}}$$\rightarrow$$
g_{s}S_{0}/2\pi l_{B}^{2}$, where $g_{s}$=2 for the spin degeneracy, we
obtain the real (absorption) part of the longitudinal magneto-optical
conductivity as
\begin{equation}
\frac{\text{Re}\sigma _{\mu \mu }}{\sigma _{0}}=\sum_{n,n^{\prime
},s,s^{\prime }}\frac{[f(E_{n^{\prime },s^{\prime }})-f(E_{n,s})]|\mu
_{n^{\prime },n}^{s^{\prime },s}|^{2}\Gamma }{(E_{n,s}-E_{n^{\prime
},s^{\prime }})[(E_{n,s}-E_{n^{\prime },s^{\prime }}+\hbar \omega
)^{2}+\Gamma ^{2}]},
\end{equation}%
where $\mu $=($x,y$), $x_{n^{\prime },n}^{s^{\prime },s}$=$\hbar
X_{n^{\prime },n}^{s^{\prime },s}/l_{B}$, $y_{n^{\prime },n}^{s^{\prime },s}$%
=$\hbar Y_{n^{\prime },n}^{s^{\prime },s}/l_{B}$, and $\sigma _{0}$=$%
2e^{2}/h $.

\section{Results and discussion}

In this section, we present the numerical results and discussions for the
magneto-optical conductivities. Hereafter, unless explicitly specified, the
conductivities are all in units of $\sigma_0=2e^2/h$, temperature $T=1$K,
Fermi energy $E_F = 0$ for interband transitions, and level broadening $%
\Gamma$=0.05$\sqrt{B}$ in unit of meV.

In order to understand the magneto-optical absorption spectra, we firstly
examine the interband magneto-optical selection rules determined by the
matrix elements of the transition rate. Figs. 3(a) and 3(b) plot all the
nonzero matrix elements of the transition rate for interband transition as a
function of the photon energy. As shown in Figs. 3(a) and 3(b), the matrix
elements $T_{x}^{vn,cn^{\prime }}$ ($T_{y}^{vn,cn^{\prime }}$) are nonzero
only if the level indexes satisfy $|n-n^{\prime }|$=0,2 ($|n-n^{\prime }|$%
=2,4), which indicates that the interband magneto-optical selection rule for
linearly polarized light along the $x$($y$)-direction is $\Delta n=0,\pm
2 $ ($\pm 2,\pm 4$), where we have defined $\Delta n=n-n^{\prime }$.
Surprisingly, the interband magneto-optical selection rule of semi-Dirac
electron system is anisotropic. This is quite different from the dipole
transition ($\Delta n$=$\pm 1$) in conventional semiconductors \cite{Laura}
and Dirac materials \cite{Ando,Tabert,Lizhou,Gusynin,Chizhova}. It also differs
from the isotropic selection rule in black phosphorus thin film \cite%
{xyzhou1,yjjiang} in spite of the highly anisotropic dispersion therein.
We have schematically illustrated the interband magneto-optical selection
rules in Fig. 3(c), where the magenta, orange and purple arrows denote the
interband transitions of $\Delta n=0,\pm 2,\pm 4$, respectively. Further, in
low photon energy regime, there are well-resolved two-peak structures in the
transition rate spectra arising from the two kinds of transitions, i.e., $%
\Delta n=0,\pm 2$ or $\Delta n=\pm 2,\pm 4$. However, the two peaks in the
transition rate spectra tend to coincide with each other with increasing
photon energy and finally merge together in high photon energy regime [see
the purple and orange lines in Fig. 3(a)]. This is actually a reflection of
the decreasing spacings in the LL spectrum plotted in Fig. 1(d). In particular, we plot the photon energies of
the three kinds of allowed interband transitions $\Delta n=0,\pm 2,\pm 4$ as
a function of level index in Fig. 3(d). As depicted in the figure, in low photon energy
regime, only the lower LLs participate in the optical transitions. There is a
large energy difference among the allowed transitions, which leads to
separated peaks in the transition rate spectra [see Figs. 3(a) and 3(b)]. However,
with the increase of photon energy, LLs with high index are involved in
the optical transitions. The difference of the nearest resonance energies
corresponding to the allowed transitions becomes smaller and smaller and
finally fades away [see Fig. 3(d)] with the increase of the level index
arising from the smaller LL spacings depicted in Fig. 1(d). This contributes
to merged peaks in the transition rate spectra in high photon energy regime
[see Figs. 3(a) and 3(b)]. Meanwhile, the transition rate shows strong
anisotropy originating from the anisotropy of the LLs, i.e, the velocity operators and the wavefunctions. The $T_{x}^{vn,cn^{\prime }}$ is two orders of magnitude larger
than $T_{y}^{vn,cn^{\prime }}$, resulting from the highly anisotropic velocity operators, which can also be inferred from the probability distributions plotted in Fig. 2.

With the help of the magneto-optical selection rule, now we can understand
the magneto-absorption spectra easier. Fig. 3(e) presents the real part of
the interband longitudinal magneto-optical conductivity as a function of
photon energy under magnetic field $B$=30 T. As shown in Fig. 3(e), the
resonance frequency of the conductivity peak varies from the mid-infrared to the
far-infrared regime for $B$=30 T. Of course, the resonance frequencies can be
modulated by varying the magnetic fields. Further, the interband magneto-optical
absorption spectra exhibit strong anisotropy inherited from the highly
anisotropic transition rate spectra. The Re$\sigma_{xx}$ (red line) is
hundreds times larger than Re$\sigma_{yy}$ (blue line) resulting from the highly
anisotropic band structure in the absence of magnetic field, i.e., the highly anisotropic velocities along different directions [see Eq. (10)].
Owing to the anisotropic magneto-optical selection rule, i.e., $\Delta
n=0,\pm 2$ ($\Delta n$=$\pm 2,\pm 4$) for linearly polarized light along $x(y)$%
-direction, the conductivity peaks in Re$\sigma_{xx}$ do not exactly
coincide with those in Re$\sigma_{yy}$. In low photon energy regime, we find
well-resolved two-peak structures in Re$\sigma_{xx}$ (Re$\sigma_{yy}$)
corresponding to the transitions of $\Delta n=0,\pm 2$ ($\Delta n$=$\pm 2,\pm
4 $). With the increase of photon energy, LLs with high index are involved
in the transition process. The differences of the nearest resonance energies
corresponding to the allowed transitions ($\Delta n$=$0,\pm 2,\pm 4$) become
smaller and smaller. They finally vanish in high photon energy regime
[see Fig. 3(d)] resulting from the decreasing LL spacing plotted in Fig.
1(d). Therefore, although the selection rule hasn't changed, we can only
find one conductivity peak in both Re$\sigma_{xx}$ and Re$\sigma_{yy}$ in high
photon energy regime. Quantitatively, the first conductivity peak in Re$%
\sigma_{xx}$ is one order of magnitude larger than others indicating a
strong absorption from $E_{v0}$ to $E_{c0}$. Other conductivity peaks in Re$\sigma_{xx}$
contributed by $\Delta n$=0 are of the same order to those contributed by $%
\Delta n$=$\pm 2$. In contrast, the conductivity peaks in Re$\sigma_{yy}$ contributed by $\Delta n$=$\pm4$
are much smaller than those contributed by $\Delta n$=$\pm 2$, which is also in
line with the transition rates shown in Fig. 3(b). Hence, Re$\sigma_{yy}$ is
dominated by the transition of $\Delta n=\pm 2$.

\begin{figure}[tbp]
\center
\includegraphics[bb=13 10 742 558, width=0.49\textwidth]{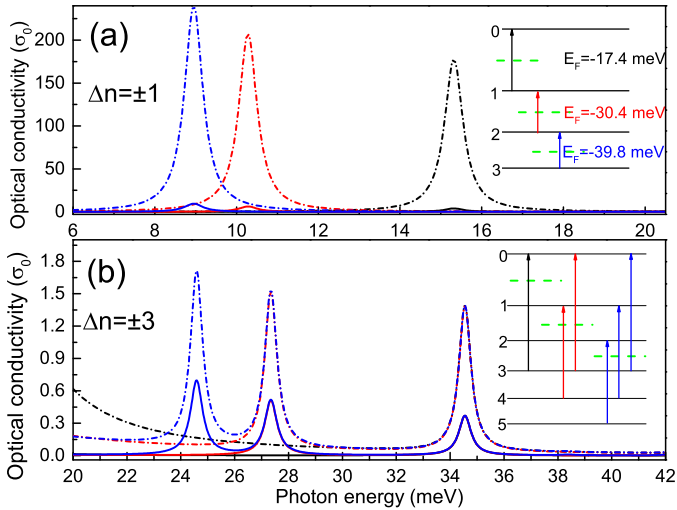} %
\renewcommand{\figurename}{Fig.}
\caption{The real part of the longitudinal magneto-optical conductivity as a
function of the photon energy under magnetic field $B$=30 T for intraband
transitions of (a) $\Delta n$=$\pm 1$ and (b) $\Delta n$=$\pm 3$,
respectively. The dash-dotted (solid) lines denote the results for polarized
light along the $x(y)$-direction. The insets depict the transitions between
the nearest LLs for transitions with filling factor $\protect\nu$=1 to 3.
The color of the arrows is the same as that of the corresponding
conductivity peaks.}
\end{figure}

Next, we turn to discuss the intraband transitions. Fig. 4 presents the real
part of the longitudinal magneto-optical conductivity as a function of the
photon energy under magnetic field $B$=30 T for intraband transitions with
filling factor $\nu$=1 to 3. The dash-dotted (solid) lines denote the
results for linearly polarized light along the $x(y)$-direction. In both Re$%
\sigma_{xx}$ and Re$\sigma_{yy}$, the intraband transitions occur when the
level index changes as $\Delta n=\pm 1, \pm 3$, contributing to two groups of
absorption peaks. This indicates that the magneto-optical selection rule of
intraband transitions is independent of the direction of polarization of
light, i.e, isotropic selection rules. This is the same as
the magneto-optical selection rules for intraband transitions in black phosphorus thin films
\cite{xyzhou1}. All the absorption peaks occur at the terahertz (THz)
frequencies. We have schematically depicted the selection rules in the
insets as the filling factor $\nu$ varying from 1 to 3. The color of the
arrows in the insets is the same as that of the corresponding conductivity
peaks. Although the selection rule is isotropic, the
magneto-optical conductivity is still highly anisotropic. Re$\sigma_{xx}$ is
one order of magnitude larger than Re$\sigma_{yy}$ because of the
anisotropic velocity operator arising from the anisotropic dispersion at zero magnetic field.
For certain Fermi level, the conductivity in both Re$\sigma_{xx}$ and Re$%
\sigma_{yy}$ contributed by the transition of $\Delta n$=$\pm 3$ is much
smaller than that contributed by $\Delta n$=$\pm 1$. Therefore, the
intraband conductivity is dominated by the dipole-type transitions ($\Delta
n=\pm 1$). Under a fixed magnetic field, the resonant frequencies for the
transitions of both $\Delta n$=$\pm 1$ and $\Delta n$=$\pm 3$ are red-shifted
with the increase of filling factor (doping), which is a direct reflection of the
decreasing LL spacings [see Fig. 1(d)]. The red-shift for the transitions of $\Delta n=\pm 3$ result in three-peak structures in the magneto-optical conductivity, which is more similar to the multi-peak structures in graphene \cite{Ando,Gusynin,Chizhova} and silicene \cite{Tabert} rather than the single peak structure in conventional semiconductors \cite{Laura}. Moreover, the
red-shift decreases with the magnetic field which can be understood from Eq.
(8). Further, we would like to point out that the interband and intraband magneto-optical conductivities reported here can be directly measured through the
infrared spectroscopy \cite{Sadowski,Jiang} or the magneto-absorption
experiments \cite{Crassee}.

\begin{figure}[t]
\center
\includegraphics[bb=18 30 671 549, width=0.49\textwidth]{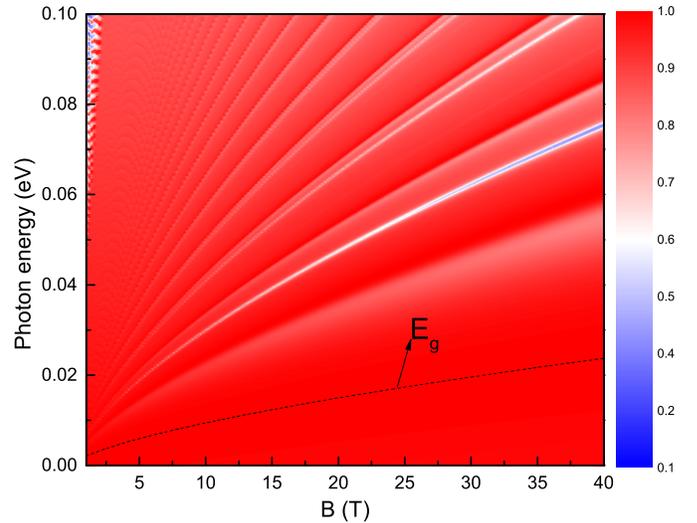} %
\renewcommand{\figurename}{Fig.}
\caption{Contour plot of the linear dichroism for interband transition as functions of photon energy and magnetic field. The black dashed line
is the band gap of LL spectrum, which corresponds to the resonance energy of
the perfect linear dichroism.}
\end{figure}

As discussed above, both the interband and intraband magneto-optical
absorption spectra are highly anisotropic, which will result in a strong
linear dichroism \cite{shuyunzhou,Bentman,Roth}. We define a dimensionless parameter $I$=$(\mathrm{Re}\sigma
_{xx}-\mathrm{Re}\sigma _{yy})/(\mathrm{Re}\sigma_{xx}+\mathrm{Re}\sigma
_{yy})$ to indicate the linear dichroism quantitatively \cite{shuyunzhou}. Fig. 5 presents a
contour plot of the linear dichroism as a function of the photon energy and
the magnetic field for the interband transitions. From the figure, we find
that $I$ is larger than 0.8 within most photon energies and magnetic fields
because Re$\sigma_{xx}$ is always dozens times larger than Re%
$\sigma_{yy}$. In principle, owing to the anisotropic selection rules, there
should be a perfect linear dichroism for the photon energy corresponding to
the transition of $\Delta n$=0 ($\Delta n$=$\pm 4$) which is only allowed in
Re$\sigma_{xx}$ (Re$\sigma_{yy}$). However, the differences of the resonance
photon energy between the transitions $\Delta n$=0 and $\Delta n$=$\pm 4$
are quenched in high photon energy regime [see Fig. 3(d)] where the perfect
linear dichroism is destroyed. Fortunately, the perfect linear dichroism
survives in low photon energy regime, where $I$ is always 1 resulting from
the transition from $E_{v0}$ to $E_{c0}$, which only can be excited by linearly
polarized along the $x$-direction. Importantly, the resonance energy of the
perfect linear dichroism is exactly the band gap of LL spectrum, which can
be effectively tunable by the magnetic field (see the black dashed line).
Therefore, we can realize a perfect linear dichroism in 2D semi-Dirac materials
with a magnetic field tunable wavelength by using the transition from $E_{v0}
$ to $E_{c0}$, which is important to design new magneto-optical devices.
There is also a strong linear dichroism for intraband transition of $\Delta n
$=$\pm 1$. It is similar to that of interband transition in the high photon
energy regime, and we do not present it here.

\section{Summary}

We have examined the LLs and magneto-optical absorption properties of a 2D
semi-Dirac electron system based on an
effective \textbf{\emph{k$\cdot$p}} Hamiltonian and linear-response theory. We found that the LLs of 2D semi-Dirac electron system can be understood as a
hybridization of those of the Schr\"{o}dinger and Dirac electron but with
new features. By using the Kubo formula, we found that the selection
rules for interband magneto-optical transitions are anisotropic with $\Delta n$=$0,\pm 2$ ($\Delta n$=$\pm 2,\pm 4$) for linearly polarized light
along the $x(y)$-direction. Whereas, the selection
rules for intraband magneto-optical transitions are $\Delta n$=$\pm 1,\pm 3$ regardless of the polarization direction of
light. For the interband (intraband) transition, the optical conductivity
for linearly polarized light along the $x$-direction is two (one) orders of
magnitude larger than that along the $y$-direction. The highly anisotropic
magneto-optical absorption spectra clearly reflect the structure of the LLs and result in strong linear dichroism. The
interband transition from $E_{v0}$ to $E_{c0}$ can realize a perfect
linear dichroism with a magnetic field tunable wavelength. The magneto-absorption spectra occur at the infrared frequency and can be detected directly by the infrared spectroscopy \cite{Sadowski,Jiang,Crassee}. Our results shed
light on the magneto-optical properties of 2D semi-Dirac electron systems and
pave the way to design magneto-optical devices based on it.

\begin{acknowledgments}
This work was supported by the National Natural Science Foundation of China
(Grant Nos. 11804092 and 11774085), Project funded by China Postdoctoral
Science Foundation (Grant Nos. BX20180097, 2019M652777), and Hunan
Provincial Natural Science Foundation of China (Grant No. 2019JJ40187).
\end{acknowledgments}


\begin{references}
\bibitem{Neto} A. H. C. Neto, F. Guinea, N. M. R. Peres, K. S. Novoselov, and A. K. Geim, \textcolor{blue} {Rev. Mod. Phys. {\bf 81}, 109 (2009)}.

\bibitem{Armitage} N.P. Armitage, E.J. Mele, and Ashvin Vishwanath \textcolor{blue} {Rev. Mod. Phys. {\bf 90}, 015001 (2018).}




\bibitem{Dietl} P. Dietl, F. Pi$\rm \acute{e}$chon, and G. Montambaux, \textcolor{blue} {Phys. Rev. Lett. {\bf 100}, 236405 (2008)}.

\bibitem{Montambaux1} G. Montambaux, F. Pi$\rm \acute{e}$chon, J.-N. Fuchs, and M. O. Goerbig, \textcolor{blue} {Phys. Rev. B {\bf 80}, 153412 (2009)}.

\bibitem{Banerjee1} S. Banerjee, R. R. P. Singh, V. Pardo, and W. E. Pickett, \textcolor{blue} {Phys. Rev. Lett. {\bf103}, 016402 (2009).}

\bibitem{Pardo1} V. Pardo and W. E. Pickett, \textcolor{blue} {Phys. Rev. Lett. {\bf102}, 166803 (2009).}

\bibitem{Huang} H. Huang, Z. Liu, H. Zhang, W. Duan, and D. Vanderbilt, \textcolor{blue} {Phys. Rev. B {\bf92}, 161115(R) (2015).}

\bibitem{Baik} S. S. Baik, K. S. Kim, Y. Yi, and H. J. Choi, \textcolor{blue} {Nano Lett. {\bf 15}, 7788 (2015).}
\bibitem{Liu} Q. Liu, X. Zhang, L. B. Abdalla, A. Fazzio, and A. Zunger, \textcolor{blue} {Nano Lett. {\bf 15}, 1222 (2015).}
\bibitem{Ghosh} B. Ghosh, B. Singh, R. Prasad, and A. Agarwal, \textcolor{blue} {Phys. Rev. B {\bf 94}, 205426 (2016)}.
\bibitem{Doh} H. Doh and H. J. Choi, \textcolor{blue} {2D Mater. {\bf 4}, 025071 (2017)}.
\bibitem{Wang1} C. Wang, Q. Xia, Y. Nie, and G. Guo, \textcolor{blue} {J. Appl. Phys {\bf 117}, 124302 (2015).}

\bibitem{Kim} J. Kim, S. S. Baik, S. H. Ryu, Y. Sohn, S. Park, B. G. Park, J. Denlinger, Y. Yi, H. J. Choi, and K. S. Kim,
\textcolor{blue} {Science {\bf 349}, 723-726 (2015).}

\bibitem{Makino} T. Makino, Y. Katagiri, C. Ohata, K. Nomura and J. Haruyama, \textcolor{blue} {RSC Adv. {\bf7}, 23427 (2017).}
\bibitem{Jang} J. Jang, S. Ahn, and H. Min, \textcolor{blue} {2D Mater. {\bf 6}, 025029 (2019)}.
\bibitem{Carbotte} J. P. Carbotte, K. R. Bryenton, and E. J. Nicol, \textcolor{blue} {Phys. Rev. B {\bf99}, 115406 (2019).}
\bibitem{Banerjee2} S. Banerjee and W. E. Pickett, \textcolor{blue} {Phys. Rev. B {\bf86}, 075124 (2012).}
\bibitem{Zhai2} F. Zhai, and J. Wang, \textcolor{blue} {J. Appl. Phys {\bf 116}, 063704 (2014).}
\bibitem{XiaoyingZhou} Wang Chen, Xianzhe Zhu, Xiaoying Zhou, and Guanghui Zhou, \textcolor{blue} {Phys. Rev. B {\bf103}, 125429 (2021).}
\bibitem{Stern} Frank Stern and W. E. Howard, \textcolor{blue} {Phys. Rev. {\bf 163}, 816 (1967).}

\bibitem{Klitzing} K. von Klitzing, \textcolor{blue} {Rev. Mod. Phys. {\bf 58}, 519 (1986)}.

\bibitem{KSNovoselov}  K. S. Novoselov, A. K. Geim, S. V. Morozov, D. Jiang, M. I. Katsnelson, I. V. Grigorieva, S. V. Dubonos and A. A. Firsov, \textcolor{blue} {Nature {\bf438}, 197 (2005).}
\bibitem{Zhangyuanbo}  Yuanbo Zhang, Yan-Wen Tan, Horst L. Stormer and Philip Kim, \textcolor{blue} {Nature {\bf438}, 201 (2005).}
\bibitem{Sadowski} M. L. Sadowski, G. Martinez, and M. Potemski, C. Berger and W. A. de Heer, \textcolor{blue} {Phys. Rev. Lett. {\bf97}, 266405 (2006).}


\bibitem{Jiang} Z. Jiang, E. A. Henriksen, L. C. Tung, Y.-J. Wang, M. E. Schwartz, M. Y. Han, P. Kim, and H. L. Stormer, \textcolor{blue} {Phys. Rev. Lett. {\bf98}, 197403 (2007).}

\bibitem{Crassee}  I. Crassee, J. Levallois, A. L. Walter, M. Ostler, A. Bost-wick, E.
Rotenberg, T. Seyller, D. Van Der Marel, and A. B. Kuzmenko, \textcolor{blue} {Nat. Phys. {\bf7}, 48 (2011).}

\bibitem{PengCheng} Peng Cheng, Canli Song, Tong Zhang, Yanyi Zhang, Yilin Wang, Jin-Feng Jia, Jing Wang, Yayu Wang, Bang-Fen Zhu, Xi Chen, Xucun Ma, Ke He, Lili Wang, Xi Dai, Zhong Fang, Xincheng Xie, Xiao-Liang Qi, Chao-Xing Liu, Shou-Cheng Zhang, and Qi-Kun Xue, \textcolor{blue} {Phys. Rev. Lett. {\bf105}, 076801 (2010).}

\bibitem{Hanaguri} T. Hanaguri, K. Igarashi, M. Kawamura, H. Takagi, and T. Sasagawa, \textcolor{blue} {Phys. Rev. B {\bf82}, 081305(R) (2010).}

\bibitem{likaiLi}Likai Li, Fangyuan Yang, Guo Jun Ye, Zuocheng Zhang, Zengwei Zhu, Wenkai Lou,
Xiaoying Zhou, Liang Li, Kenji Watanabe, Takashi Taniguchi, Kai Chang, Yayu Wang,
Xian Hui Chen and Yuanbo Zhang, \textcolor{blue}{Nat. Nanotech. {\bf 11}, 593 (2016)}.

\bibitem{xyzhou1} Xiaoying Zhou, Wen-Kai Lou, Feng Zhai and Kai Chang, \textcolor{blue} {Phys. Rev. B {\bf92}, 165405 (2015).}

\bibitem{xyzhou2} X. Y. Zhou, R. Zhang, J. P. Sun, Y. L. Zou, D. Zhang, W. K. Lou, F. Cheng, G. H. Zhou, F.
Zhai and Kai Chang, \textcolor{blue} {Sci. Rep. {\bf5}, 12295 (2015).}

\bibitem{xyzhou3} Xiaoying Zhou, Wen-Kai Lou, Dong Zhang, Fang Cheng, Guanghui Zhou, and Kai Chang, \textcolor{blue} {Phys. Rev. B {\bf95}, 045408 (2017).}

\bibitem{yjjiang} Yongjin Jiang, Rafael Rold\'{a}n, Francisco Guinea, and Tony Low, \textcolor{blue} {Phys. Rev. B {\bf92}, 085408 (2015).}


\bibitem{Smith} G. D. Smith, \emph{Numerical Solutions of Partial Differential Equations: Finite Difference Methods} (Oxford Univ. Press, Oxford, 1978), 2nd ed.



\bibitem{TAndo1} T. Ando and Y. Uemura, \textcolor{blue} {J. Phys. Soc. Japan {\bf 36}, 959 (1974).}

\bibitem{Ando} Mikito Koshino and Tsuneya Ando, \textcolor{blue} {Phys. Rev. B {\bf 77}, 115313 (2008).}


\bibitem{Tabert} C. J. Tabert and E. J. Nicol, \textcolor{blue} {Phys. Rev. Lett. {\bf 110}, 197402 (2013).}

\bibitem{Lizhou} Zhou Li and J. P. Carbotte, \textcolor{blue} {Phys. Rev. B {\bf 88}, 045414 (2013).}



\bibitem{Esfarjani} K. Esfarjani, H. R. Glyde, and V. Sa-yakanit, \textcolor{blue} {Phys. Rev. B {\bf 41}, 1042 (1990).}

\bibitem{Sari} J. S\'{a}ri, M. O. Goerbig, and Csaba T\H{o}ke, \textcolor{blue} {Phys. Rev. B {\bf 92}, 035306 (2015).}

\bibitem{Zettili} N. Zettili, Quantum Mechanics: Concepts and Applications, 2nd ed. (Wiley, 2009).

\bibitem{Laura} Laura M. Roth, Benjamin Lax, and Solomon Zwerdling, \textcolor{blue} {Phys. Rev. {\bf 114}, 90 (1959)}.
\bibitem{Chizhova} L. A. Chizhova, J. Burgd\"{o}rfer, and F. Libisch, \textcolor{blue} {Phys. Rev. B {\bf92}, 125411 (2015).}

\bibitem{Gusynin} V. P. Gusynin, S. G. Sharapov, and J. P. Carbotte, \textcolor{blue} {Phys. Rev. Lett. {\bf 98}, 157402 (2007).}

\bibitem{shuyunzhou}Changhua Bao, Hongyun Zhang, Teng Zhang, Xi Wu, Laipeng Luo, Shaohua Zhou, Qian Li, Yanhui Hou, Wei Yao, Liwei Liu, Pu Yu, Jia Li, Wenhui Duan, Hong Yao, Yeliang Wang, and Shuyun Zhou,
\textcolor{blue} {Phys. Rev. Lett. {\bf 126}, 206804 (2021)}.

\bibitem{Bentman} H. Bentmann, H. Maa{\ss}, E. E. Krasovskii, T.R.F. Peixoto, C. Seibel, M. Leandersson, T. Balasubramanian, and F. Reinert,
\textcolor{blue} {Phys. Rev. Lett. {\bf 119}, 106401 (2017)}.

\bibitem{Roth} Ch. Roth, F. U. Hillebrecht, H. B. Rose, and E. Kisker, \textcolor{blue} {Phys. Rev. Lett. {\bf 70}, 3479 (1993)}.

\end{references}
\end{document}